%% file: main.tex
\pdfoutput=1
\documentclass[10pt,conference]{IEEEtran}
\IEEEoverridecommandlockouts

\usepackage{cite}
\usepackage{amsmath,amssymb,amsfonts}
\usepackage{graphicx}
\usepackage{textcomp}
\usepackage{xcolor}
\usepackage{booktabs}
\usepackage{balance}
\usepackage[hyphens]{url}
\usepackage[hidelinks]{hyperref}
\hypersetup{pdftitle={MVP: A Motion-Predictive Speculative Vision Pipeline with Non-Blocking Drift Correction},pdfauthor={Raul Taranco and Antonio Gonz\'alez}}
\usepackage[tracking=true]{microtype}
\usepackage{orcidlink}
\usepackage{eso-pic}
\AddToShipoutPictureFG*{%
  \AtPageLowerLeft{\put(\LenToUnit{0.75in},\LenToUnit{0.3in}){\parbox{7in}{\scriptsize\centering
    \copyright~2026 IEEE. Personal use of this material is permitted. Permission from IEEE must be obtained for all other uses, in any current or future media, including reprinting/republishing this material for advertising or promotional purposes, creating new collective works, for resale or redistribution to servers or lists, or reuse of any copyrighted component of this work in other works.}}}}

\newcommand{\revised}[1]{#1}

\input{tex/0_vars}

\title{\name{}: A Motion-Predictive Speculative Vision Pipeline with Non-Blocking Drift Correction}

\author{%
  \IEEEauthorblockN{Raul Taranco\,\orcidlink{0000-0002-1564-4365}}
  \IEEEauthorblockA{\textit{Universitat Polit\`ecnica de Catalunya}\\
    Barcelona, Spain\\
    raul.taranco@upc.edu}
  \and
  \IEEEauthorblockN{Antonio Gonz\'alez\,\orcidlink{0000-0002-0009-0996}}
  \IEEEauthorblockA{\textit{Universitat Polit\`ecnica de Catalunya}\\
    Barcelona, Spain\\
    antonio@ac.upc.edu}
}

\begin{document}
\maketitle

\begin{abstract}

Continuous Vision (CV) systems underpin real-time applications such as autonomous driving and augmented reality, where latency, throughput, and energy are tightly constrained on mobile platforms. Modern CV SoC pipelines, however, still serialize image capture and processing, leading to high end-to-end latency. Prior work reduces this latency by predicting future frames and running pixel-domain backend inference speculatively, but incorrect predictions force re-execution on real frames, increasing energy and complexity.

We present MVP, a motion-predictive speculative vision pipeline that operates entirely in the motion domain. Instead of forecasting full images, MVP predicts future motion vectors and uses them to extrapolate perception results from previously processed frames before the next frame arrives. A lightweight hardware extension in the Image Signal Processor (ISP) reuses existing motion-estimation logic to predict motion with minimal area and energy cost.

MVP introduces a scheduling model that treats motion extrapolation as the default path, while full backend inference runs periodically in the background for drift correction off the critical path. It also supports optional frontend scaling, allowing the system to reduce sensor sampling under low or predictable motion to save energy.

We evaluate MVP on object detection, demonstrating up to 66.8\% reduction in tail latency and 46\% energy savings, at a small accuracy cost.

\end{abstract}

\input{tex/1_introduction}

\input{tex/2_background}

\input{tex/3_proposal}
\input{tex/4_arch_proposal}
\input{tex/5_methodology}
\input{tex/6_results}
\input{tex/7_related_work}
\input{tex/8_conclusions}

\section*{Acknowledgments}
This work has been supported by the Spanish State Research Agency (MCIU/AEI/10.13039/501100011033), FEDER/UE and the European Social Fund Plus (ESF+) under grants PID2020-113172RB-I00 and PID2024-155476OB-I00, and the ICREA Academia program.

\IEEEtriggercmd{\balance}
\IEEEtriggeratref{20}
\bibliographystyle{IEEEtranS}
\bibliography{refs}

\end{document}

%% file: tex/0_vars.tex
\newcommand{\name}{MVP}

\newcommand{\latencyreductionOD}{66.8\%}  %
\newcommand{\energyreductionOD}{46\%}

%% file: tex/1_introduction.tex
\section{Introduction}\label{sec:introduction}

Continuous Vision (CV) systems are critical enablers of advanced machine perception, driving progress across diverse fields such as Autonomous Driving (AD)~\cite{yurtseverSurveyAutonomousDriving2020}, Augmented/Virtual/Extended Reality (AR/VR/XR)~\cite{hamadHowVirtualReality2022,chatzopoulosMobileAugmentedReality2017,silvaEXtendedRealityXR2022}, robotics, and smart vision systems~\cite{alyamkinLowPowerComputerVision2019}. However, their deployment on mobile and resource-constrained platforms reveals a fundamental tension between real-time performance, high-throughput processing, and limited energy budgets. This tension is exacerbated by the rising demand for high-resolution, high-frame-rate camera sensors~\cite{ishii2000FpsRealtime2010,galoogahiNeedSpeedBenchmark2017,nishimura8K4Kresolution60fps450ke2018}.

In safety-critical domains like AD, even minimal delays can be hazardous, particularly at high speeds~\cite{davidCAR2XPedestrianSafety2010,hsiaoZhuyiPerceptionProcessing2022a}. In AR/VR/XR scenarios, latency can lead to user discomfort and degrade the overall immersive experience~\cite{duzmanskaCanSimulatorSickness2018}. These applications are particularly sensitive to timing constraints, often requiring end-to-end latencies below 100 milliseconds for AD and under 50 milliseconds for XR to maintain responsiveness and reliability~\cite{davidCAR2XPedestrianSafety2010,duzmanskaCanSimulatorSickness2018}. At the same time, energy efficiency remains a principal concern on mobile devices because shorter battery lifetimes limit operational autonomy.

Modern CV SoCs implement a pipelined frontend-backend architecture, where a camera sensor and ISP (frontend) feed a perception backend (e.g., CPU, NPU), as detailed in Section~\ref{sec:cv_bck}. Although the frontend and backend can run concurrently on different frames in a pipelined manner, the latency for any single frame remains constrained by the sequential handoff between sensing and processing.

Prior work addresses these bottlenecks from two directions (see Section~\ref{sec:prior_approaches}). Speculative execution, introduced in PVF~\cite{ganLowLatencyProactiveContinuous2020}, predicts entire future frames and runs backend inference ahead of time, validating speculative results when actual frames arrive. While this can hide inference latency, it requires full-frame pixel synthesis, additional compute resources (e.g., NPUs/GPUs), and suffers from high misprediction penalties and periods without fresh sensor data when the frontend is disabled to save energy. Motion extrapolation, exemplified by Euphrates~\cite{zhuEuphratesAlgorithmSoCCodesign2018}, reduces the frequency of full backend processing by translating previously computed perception outputs forward in time using ISP-generated motion vectors. This approach substantially lowers energy and average latency but key frames still incur full backend latency, and extrapolated results may be delayed or disrupted when frames are dropped.

To address the previous limitations, we introduce \name{}, the first speculative pipeline that unifies vision speculation and motion extrapolation by forecasting future motion vectors instead of full frames as in prior work. The novelty of \name{} lies not in speculation or extrapolation individually, but in their integration entirely within the motion domain. This design eliminates pixel synthesis, specialized inference engines, and the high penalties of misprediction, making \name{} considerably more efficient and easier to deploy on commodity CV SoCs.

Specifically, \name{} introduces three core innovations: (1) a decoupled drift control scheduling model that proactively applies motion extrapolation, (2) motion vector prediction enabling frame-level speculation before actual frame capture, and (3) optional frontend scaling to reduce sensing energy under stable scene conditions.

Operating solely in the motion domain greatly reduces complexity. Modern ISPs already generate motion vectors at $8\times8$ or $16\times16$ pixel granularities, thus eliminating the need for additional computational overhead. \name{} leverages this readily available data for efficient, proactive speculation.

Unlike Euphrates, \name{} treats motion extrapolation as a first-class scheduling primitive, proactively applying it immediately to every incoming frame to minimize latency. Full backend inference runs periodically and asynchronously in the background to correct any drift accumulated over multiple extrapolated frames. The frequency of drift corrections directly controls the trade-off between accuracy and energy savings. When using extrapolation intervals comparable to Euphrates, \name{} achieves similar accuracy but with much lower tail latency.

Proactive motion extrapolation alone, however, does not eliminate the latency associated with capturing the frame itself. Therefore, \name{} further introduces speculative prediction of motion vectors for future frames, enabling speculative perception results before the actual frame arrives. Upon frame capture, the predicted vectors are rapidly validated against the actual ISP-generated vectors. Only when discrepancies exceed a threshold in critical image regions, for instance, around detected objects, the system triggers a fast, targeted correction step.

\name{} also enables optional frontend scaling based on scene dynamics. When predicted motion is low and stable, the system may reduce sensor temporal resolution to conserve energy while still collecting sufficient data to refine motion prediction. Unlike PVF, and thanks to the low overhead of speculation, this optimization is optional and can be selectively applied to reduce energy consumption without compromising accuracy.%

To support these capabilities we integrate a lightweight hardware unit into the ISP that builds on the motion estimation already performed by typical ISPs. This unit tracks motion vectors for each image block over time and predicts future motion vectors using temporal modeling. Since consecutive frames in continuous vision workloads are often similar, predicting motion a few frames ahead becomes feasible with simple methods. We use a second-order autoregressive predictor, inspired by classic object tracking models, which estimates future motion vectors as a linear combination of past observations for each image block. While more complex or machine learning-based predictors are possible, this simple approach performs well in practice. Additionally, by accelerating it in hardware, we enable multiple processing units to benefit independently and amortize our new proposed functionality. %

We evaluate \name{} using object detection as our primary workload, given its representativeness, widespread deployment, and suitability for direct comparison with prior art. \name{} consistently maintains high accuracy while substantially improving tail latency and energy efficiency. Specifically, it achieves up to \latencyreductionOD{} reduction in tail latency and \energyreductionOD{} in energy consumption, at a small accuracy cost. The same motion-domain primitives and scheduling extend directly to object tracking (propagating tracks using predicted MV fields), camera localization/SLAM~\cite{mur-artalORBSLAM2OpenSourceSLAM2017,camposORBSLAM3AccurateOpenSource2021} (extrapolating features for backend optimization), and activation–motion compensation in CNNs (e.g., EVA$^2$~\cite{bucklerEVAExploitingTemporal2018}) that update intermediate activations using motion cues rather than being recomputed.

In summary, the primary contributions of this work are:
\begin{itemize}
\item A new scheduling model that delivers immediate perception results using motion extrapolation while decoupling backend processing into a non-blocking background task, reducing the effective latency while preserving accuracy.

\item A motion-domain speculation with non-blocking drift control scheme, which avoids pixel synthesis and parallel full-inference, slashes the cost of mispredictions, and keeps integration simple (ISP-integrated predictor without dedicated NPU).

\item A lightweight hardware extension embedded in the ISP that predicts future motion vectors using second-order temporal models. This design builds on motion estimation already performed in conventional ISPs and enables low-latency speculation.

\item Experimental validation showing that \name{} substantially reduces tail latency and energy consumption, making it practical for deployment in real-time CV workloads.
\end{itemize}

%% file: tex/2_background.tex
\section{Background and Motivation}
\label{sec:motivation}

\subsection{Continuous Vision Pipeline}
\label{sec:cv_bck}

\begin{figure}[t]
    \centering
    \includegraphics[width=\linewidth]{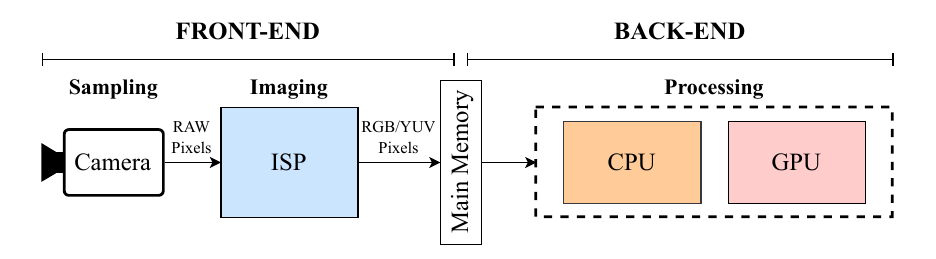}
    \caption{Overview of the CV pipeline, illustrating the interaction between the frontend (sensor and ISP) and backend.}%
    \label{fig:vision_pipeline}
\end{figure}

\begin{figure*}[th]
    \centering
    \includegraphics[width=\textwidth]{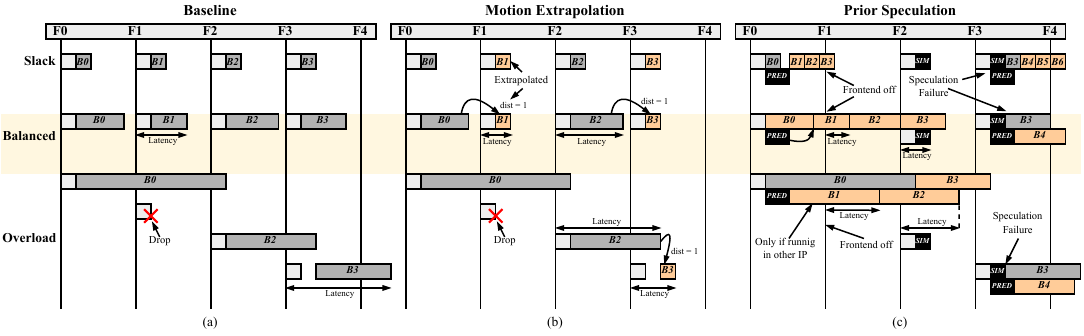}
    \caption{Comparison of frame processing strategies across three compute regimes in real-time computer vision systems.
\textit{Baseline} uses a single in-flight frame and a queue that retains only the latest input.
\textit{Motion Extrapolation} performs full processing on selected frames and extrapolates intermediate ones using motion vectors.
\textit{Speculation} predicts and processes future frames ahead of arrival (PRED), validating them upon frame arrival (SIM). \textit{Slack}, \textit{Balanced}, and \textit{Overload} regimes show how system behavior shifts as backend pressure increases. $dist$ is the extrapolation distance: number of frames between two full backend processing stages.}
    \label{fig:regimes_cv}
\end{figure*}

CV systems are always-on pipelines that continuously capture, process, and interpret visual information to drive fast, accurate responses. They consist of two main components: the frontend and the backend, as shown in Figure~\ref{fig:vision_pipeline}. The frontend captures and processes images from the environment and prepares them for further analysis or display. The off-chip camera sensor captures light and converts it into a RAW pixel image. Later, it transmits this image to the SoC through the MIPI Camera Serial Interface (CSI)~\cite{limMultilaneMIPICSI2010}. In the SoC, the Image Signal Processor (ISP) transforms the raw pixel data into a visually enhanced RGB/YUV image, usually saved in a framebuffer within the SoC's main memory. The backend derives high-level semantic information from these images, facilitating applications like object detection, tracking, or ego-localization. In practical deployments, the backend's responsibilities extend beyond just perceiving the environment. It must also produce actionable output that triggers the next stage of the system, be it navigating around an obstacle, rendering a new scene in a headset, or updating a control command in a robotic platform. In robotics and autonomous vehicles, for example, perception results feed into downstream tasks, such as planning and control layers, in frameworks like ROS~\cite{macenskiRobotOperatingSystem2022} or Autoware~\cite{Autoware}. In AR/VR/XR systems, perceived scene geometry or user pose must immediately drive rendering engines to maintain immersive realism.

These downstream stages impose strict latency constraints on the vision pipeline: perception must be not only fast, but also timely enough to support real-time decision-making.

\subsection{Latency Regimes and Flow-Control Behavior}\label{sec:regimes_cv}

We define \textit{frame latency} as the time from the start of image sensing to the completion of semantic processing. \revised{This frame latency measures how quickly a captured frame becomes an output. We separately define \textit{information-propagation latency} as the time for newly appearing scene content to first reach an output.} In real-time CV systems, latency and throughput hinge on how the pipeline manages incoming frames under computing constraints.

To ensure responsiveness and prevent unbounded latency, real-time CV systems often adopt bounded flow-control policies. These limit the number of frames processed concurrently, defined by the number of in-flight frames, and constrain the input queue to hold only the most recent unprocessed frame. The number of in-flight frames depends on the available compute resources for a given vision task; in our case, we assume a single processing unit. We also assume a queue of size one, which stores only the latest arriving frame while dropping older inputs. This standard ``process-latest, drop-oldest'' policy prioritizes freshness over completeness, reduces queuing latency, and aligns with the design of latency-sensitive CV systems~\cite{MediapipeMediapipeCalculators}. We adopt this model throughout our analysis to study how different compute regimes impact latency and throughput. In this work, we define tail latency as the 99th-percentile sensor-to-task completion time under a bounded, process-latest, drop-oldest queue of 1 unless stated otherwise.

Figure~\ref{fig:regimes_cv}a illustrates the behavior of a conventional CV pipeline across three representative compute regimes: \textit{Slack}, \textit{Balanced}, and \textit{Overload}. The diagram shows how the baseline system processes five consecutive frames (\texttt{F0-F4}) as the system load increases. Each regime is depicted as a horizontal sequence, with time progressing from left to right. Light gray segments represent frontend stages (e.g., sensing, transmission, ISP), and dark gray segments represent backend processing (e.g., object detection). Boxes labeled \texttt{BN} denote backend processing for frame \texttt{FN} (e.g., \texttt{B0} corresponds to backend processing for frame \texttt{F0}).

In the \textit{Slack regime}, backend processing comfortably completes before the next frame arrives. No frames drop, latency remains minimal, and throughput matches the camera's native frame rate. However, most latency in this regime stems from frontend processing (sensor exposure, readout, transmission to the SoC, and ISP), which makes backend latency optimizations progressively less impactful. Consequently, the frontend dominates energy consumption in this regime.

In the \textit{Balanced regime}, backend processing is tightly matched with frame arrival intervals. The processing unit consistently finishes inference around the time the next frame arrives, potentially introducing only slight or sporadic additional latency. This regime represents the most balanced operating point, achieving low latency, no dropped frames, and high backend utilization. Under these conditions, backend optimizations directly translate into noticeable improvements in system latency and energy efficiency~\cite{zhuEuphratesAlgorithmSoCCodesign2018}. %

In the \textit{Overload regime}, backend processing far exceeds the frame arrival interval. Due to the bounded flow-control strategy, the pipeline continuously overwrites the queued frame with new arrivals, resulting in frequent frame drops, as seen with frame \texttt{F1}. Although latency remains bounded (due to the dropping mechanism), throughput sharply declines, causing loss of temporal resolution and responsiveness. Demanding workloads such as high-precision object detection at 30 FPS often push mobile systems into this regime.

\subsection{Key Optimizations in Prior Work}
\label{sec:prior_approaches}

Two primary approaches, \textit{motion extrapolation} and \textit{speculative execution}, have been proposed to address the performance bottlenecks of CV systems. 

\subsubsection{Motion Extrapolation}\label{subsec:mot_extrapolation}

Motion extrapolation is a lightweight strategy for reducing the computational load in vision pipelines by avoiding full inference on every frame. Euphrates~\cite{zhuEuphratesAlgorithmSoCCodesign2018} is a representative implementation that introduces a software-hardware co-design to make this technique practical for mobile CV systems. Instead of processing every frame, Euphrates performs high-precision full inference on a subset of frames (I-frames) and extrapolates the intermediate ones (E-frames) using motion vectors extracted from standard ISPs. 

Figure~\ref{fig:regimes_cv}b illustrates how this approach reduces the latency of the extrapolated frames (E-frames, shown in orange) and lowers the overall energy consumption. These benefits become particularly pronounced in the Balanced and Overload regimes, where reducing the frequency of computationally expensive inference substantially cuts backend demands. In contrast, in the Slack regime, where frontend latency dominates, extrapolation offers diminishing returns.

The number of consecutive E-frames between two I-frames defines the extrapolation distance. In Figure~\ref{fig:regimes_cv}b, this distance is one (dist $=$ 1), meaning every other frame is extrapolated. However, the original Euphrates evaluation explores a wider viable range of extrapolation distances, from 2 up to 32 frames, and shows that distances between 4 and 8 E-frames offer a sweet spot that incurs modest accuracy drops while sharply lowering compute cost in object detection and tracking tasks.

Nevertheless, Euphrates still faces notable limitations. In its design, I-frames still incur the full processing latency (e.g., frame \texttt{F2}), which restricts reductions in overall system-wide tail latency. Moreover, Euphrates uses a scheduling strategy in which E-frames must wait for the previous I-frame to finish before extrapolated results become available. When the backend becomes overloaded, the system may drop several frames entirely, disrupting the extrapolation sequence. This issue is evident in the Overload regime of Figure~\ref{fig:regimes_cv}b, where frame \texttt{F1} is discarded by the flow control mechanism, as in the baseline.

\subsubsection{Speculation}\label{subsec:cv_speculation}

Speculative CV execution was first introduced in PVF~\cite{ganLowLatencyProactiveContinuous2020} as an aggressive approach to reduce average latency. As illustrated in Figure~\ref{fig:regimes_cv}c, this approach allows the system to maintain high throughput even under the \textit{Overload} regime, as backend processing occurs ahead of actual frame arrivals. Instead of processing frames only after their arrival, PVF predicts (\texttt{PRED}) and fully processes multiple future frames speculatively (\texttt{B1}, \texttt{B2}, and \texttt{B3}), exploiting parallelism across different system IPs (e.g., CPU, GPU, and NPU). When an actual frame arrives (e.g., \texttt{F2}), the system compares it with predicted frames (\texttt{SIM}) using a global similarity metric. If the match is within a predefined threshold, PVF commits the speculative results, hiding the inference latency.

However, speculative models like PVF incur substantial complexity and energy overhead. They must generate high-quality, full-resolution pixel-domain predictions and perform complete backend processing for each speculative frame. They also depend on having multiple compute units available, such as NPUs for frame prediction and additional IPs for inference, which may not exist on all platforms or may be allocated to other tasks. For example, in the Overload regime of Figure~\ref{fig:regimes_cv}c, speculation is only effective because an additional processing unit is assumed to handle speculative inference for \texttt{B1} while the main unit processes \texttt{B0}. Without this extra computing resource, speculation and regular processing would serialize, providing no latency advantage. Furthermore, speculative tasks often run on less energy-efficient processors: while \texttt{B0} may execute on an NPU, \texttt{B1} may fall back to a GPU, increasing power consumption. Compounding these issues, when speculation fails (e.g., for frame \texttt{F3}), the system must reprocess the frame, wasting compute resources and increasing tail latency.

To mitigate the high energy cost, PVF periodically disables the vision frontend (e.g., \texttt{F1} frame) during speculative execution. However, this introduces ``blind spots'', periods when the system lacks access to fresh sensor data, making PVF less suitable for safety-critical applications.

These challenges highlight the need for execution models that preserve the low latency benefits of speculation, without current energy penalties and increased system complexity.

%% file: tex/3_proposal.tex
\section{\name{}: Motion-Based Speculative Vision}\label{sec:proposal_th}

\subsection{Overview and Core Insight}

\begin{figure}[t]
\centering
\includegraphics[width=0.9\linewidth]{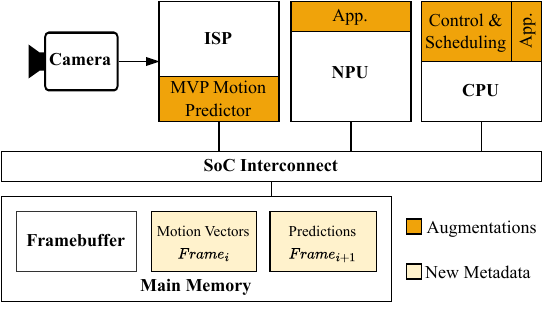}
\caption{Overview of the proposed SoC augmentations.}
\label{fig:soc}
\end{figure}

Our key insight is to reimagine the ISP not only as a passive imaging component but also as a lightweight speculative engine capable of predicting future motion. \name{} feeds the motion vectors already generated within modern ISPs to a lightweight predictor embedded in the ISP, which estimates the motion vectors of the next frame before it arrives. As shown in Figure~\ref{fig:soc}, the system stores these predictions as metadata alongside the framebuffer in main memory, making them accessible to the rest of the SoC. We expose these predictions as control signals and delegate the decision-making logic to programmable components already present in the system, such as microcontroller units (MCUs) or CPUs, which can dynamically adjust the CV pipeline based on anticipated scene dynamics. This design minimizes hardware modifications, avoids assumptions about the final application or frame scheduling, and supports flexible integration across a wide range of use cases.%

\name{} improves upon prior extrapolation schemes described in Subsection~\ref{subsec:mot_extrapolation} by introducing a new scheduling model that removes full-frame inference from the critical path. Each frame is served immediately via extrapolation, while high-precision frame processing is performed asynchronously on selected \textit{Drift-Control} (DC) frames to periodically correct prediction drift. This reduces end-to-end response time and allows perception results to arrive earlier for downstream tasks such as control or planning.

In contrast to previous speculation CV pipelines such as PVF~\cite{ganLowLatencyProactiveContinuous2020}, discussed in Subsection~\ref{subsec:cv_speculation}, \name{} does not generate full-resolution predicted frames or require running multiple inference engines in parallel to be effective. Instead, it speculates only on motion vectors, which are inexpensive to compute and validate. This design dramatically reduces the penalty of misprediction and removes the need for aggressive power-saving strategies, such as disabling the vision frontend. 

While not required by default, \name{} optionally supports motion-aware frontend scaling by adjusting capture parameters such as frame triggering based on predicted motion magnitude. This allows the system to further reduce sensing energy in low-motion or predictable scenarios, offering additional efficiency gains without compromising responsiveness.

\name{} achieves all of the above with minimal hardware overhead. It requires no additional NPUs for frame prediction, no complex scheduling logic, and no dedicated microcontrollers. The design builds on existing ISP motion vectors and introduces a lightweight hardware extension to expose motion predictions to the rest of the SoC. 

In summary, \name{} introduces three novel mechanisms:
\begin{itemize}
    \item A scheduling model based on always-on extrapolation, with \textit{Drift-Control} frames for periodic correction.
    \item Lightweight speculative execution using motion vector prediction and local validation.
    \item Adaptive frontend scaling that reduces sensing energy based on predicted motion magnitude.
\end{itemize}

\name{} can be regarded as a stepwise evolution from a motion extrapolation baseline (illustrated in Figure~\ref{fig:regimes_cv}b).

\subsection{Decoupled Scheduling via Drift-Control Frames}\label{sec:DCFD}

Earlier motion extrapolation techniques (Section~\ref{subsec:mot_extrapolation}) rely on a sequential scheduling model that tightly couples extrapolation with full-frame processing. \name{} overcomes this limitation by decoupling motion extrapolation from full-frame processing. We propose a simple scheduling model, \name{}-SCH, that treats motion extrapolation as the default processing path for every frame, producing immediate results. Full backend processing runs on selected frames, termed \textit{Drift Control} (DC) frames, which correct accumulated drift and re-align the system with ground truth. 

We define a parameter, \textit{Drift Control Frame Distance (DCFD)}, to control the frequency of these corrections. DCFD specifies the number of frames between two DC frames. When DCFD is set to zero, the system performs DC on every frame, which eliminates drift but provides no energy savings. Increasing the DCFD reduces how often the backend executes, introducing a tunable trade-off between accuracy and efficiency. \revised{We also define the \textit{Max Extrapolation Distance (MED)}, the largest number of frames the system extrapolates after a DC frame, which bounds the accumulated drift before the next correction.}

\name{} supports both fixed and adaptive DCFD tuning. In adaptive mode, the scheduler continuously monitors the accuracy of motion extrapolation by comparing the output of a DC frame with the extrapolated result for the same frame (for example, comparing \texttt{DC0} with \texttt{B0}). If the deviation stays below a predefined threshold, the scheduler increases the DCFD (up to a configurable maximum) to reduce the number of DC frames and conserve energy. When the deviation exceeds the threshold, the system lowers the DCFD to trigger more frequent corrections. If errors persist or grow, the system progressively increases the correction rate and eventually returns to baseline processing if necessary. We implement this adaptive behavior using an exponential backoff strategy that adjusts DCFD based on the magnitude and frequency of observed errors.

\begin{figure}[b]
\centering
\includegraphics[width=0.9\linewidth]{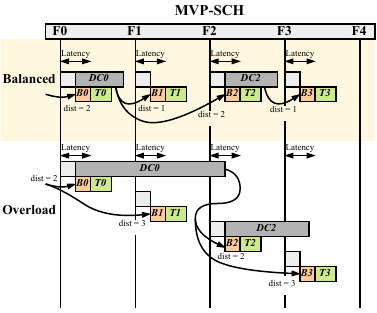}
\caption{Overview of \name{}'s new scheduling proposal. \textit{dist} accounts for the extrapolation distance (in number of frames) since the last DC execution.}%
\label{fig:mpv_sch}
\end{figure}

Figure~\ref{fig:mpv_sch} revisits the Balanced and Overload regimes from Figure~\ref{fig:regimes_cv}, applying the \name{}-SCH scheduling model with a DCFD$=$1. Each frame (\texttt{F0} to \texttt{F4}) undergoes motion extrapolation (labeled \texttt{B0} to \texttt{B3}), producing results with minimal delay. Note how frame \texttt{F1} now is not dropped since it does not need to wait until \texttt{F0} generates results to perform the extrapolation. This lightweight operation can run on a low-power CPU or microcontroller, making it suitable for mobile platforms.

Beyond perception, many CV systems include additional downstream tasks that consume perception outputs to perform other tasks such as control and planning. We denote these tasks as \texttt{T0} through \texttt{T3} in the figure. \revised{Motion extrapolation accelerates the perception stage, while these downstream tasks cannot run it themselves. They still benefit from \name{}, because \name{} produces a perception output for every frame, sooner than a conventional pipeline and even for frames that pipeline would otherwise drop under load. These tasks therefore receive timely and uninterrupted inputs.} For instance, systems may use extrapolated bounding boxes for trajectory planning~\cite{linArchitecturalImplicationsAutonomous2018} or pass visual features to SLAM optimization backends~\cite{mur-artalORBSLAM2OpenSourceSLAM2017,camposORBSLAM3AccurateOpenSource2021} to estimate the camera pose more quickly.

In summary, \name{}-SCH eliminates blocking dependencies between extrapolated and backend frames, ensuring timely perception outputs even in the Overload regime. Furthermore, it dynamically adjusts the drift correction rate based on available compute resources, minimizing energy consumption while guaranteeing the required level of precision.

\subsection{Speculative Motion Prediction} %

\name{}-SCH offers a lightweight way to generate immediate perception results for every frame. However, it cannot hide the latency introduced by motion extrapolation itself, nor the latency from downstream consumers (i.e., \texttt{T0-T3} tasks) of the perception results that cannot leverage extrapolation and still contribute to the overall system latency. To address this, we introduce \name{}-SPE: a motion-based speculative CV pipeline.

\name{}-SPE speculates directly in the motion domain. As we will see in Section~\ref{sec:arch_proposal}, the new \name{} ISP is capable of predicting the motion vectors of the upcoming frames in advance and making them available to the rest of the SoC. This allows the system to generate preliminary perception results via speculative motion extrapolation even before the frame is captured, hiding extrapolation latency and improving responsiveness for downstream consumers.

The \name{}-SPE scheduler triggers speculative motion extrapolation at two key events: when extrapolation for the current frame completes or when a DC frame finishes execution. If speculating on the current frame exceeds the MED limit, the scheduler pauses speculation to avoid excessive drift until a DC frame provides fresh results. The \textit{Max Speculation Distance (MSD)} sets the maximum number of frames in the future for which the scheduler can initiate speculation. In other words, it bounds the distance between the current frame and the farthest future frame eligible for speculative extrapolation. If the scheduler does not trigger speculation for a frame, it schedules the frame for regular processing, following \name{}-SCH behavior.

In all our experiments, we use MSD$=$1 unless stated otherwise. This setting corresponds to the minimum speculation distance and is sufficient in most cases to hide the latency of motion extrapolation and downstream tasks. While higher speculation distances are feasible, they typically provide little additional benefit and increase the risk of mispredictions.

\begin{figure}[t]
\centering
\includegraphics[width=0.9\linewidth]{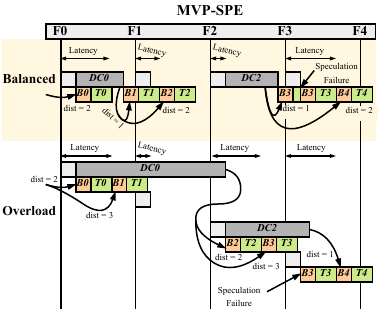}
\caption{Overview of the \name{} speculative CV pipeline with DCFD$=$1, MSD$=$1, and MED$=$2 and 3 for Balanced and Overload regimes, respectively. On MV arrival for frame $i+1$, predicted MVs are validated; mismatches trigger a local re-extrapolation.} 
\label{fig:mpv_spe}
\end{figure}

Figure~\ref{fig:mpv_spe} illustrates this speculative pipeline in action. In the Balanced regime, after receiving frame \texttt{F0}, the system performs motion extrapolation in \texttt{B0} and immediately executes the dependent task \texttt{T0}. Using the motion vectors and the output of \texttt{B0}, the system speculates on the next frame's result (\texttt{B1})  and proactively executes the next task \texttt{T1} ahead of time.

In the Balanced regime, we typically set the MED$=$DCFD$+$1 (MED$=$2 in this case) since it is sufficiently aggressive to allow effective speculation in our evaluations.  As a result, the scheduler cannot speculate immediately after \texttt{B0}, since doing so would exceed the MED limit by extrapolating three frames past the last DC frame. Instead, speculation is deferred until \texttt{DC0} completes. Once fresh results are available, the scheduler speculates on \texttt{B1}, partially hiding the latency of frame \texttt{F1}. Later, the scheduler immediately speculates on \texttt{B2}, fully hiding its latency and maintaining uninterrupted task execution. \texttt{B3} starts speculative execution after \texttt{DC2}. However, the motion vector prediction fails and the scheduler re-extrapolates \texttt{B3}, reverting to the \name{}-SCH behavior. Importantly, because this speculation operates in the motion domain, the misprediction penalty is much lower than in prior pixel-domain methods. %

In the Overload regime, the system behaves similarly but requires a higher MED value (MED$=$3 in this case). Without this adjustment, speculation would not trigger due to the longer delay in DC frame completion. Notably, a MED$=$3 matches the extrapolation distance already used in \name{}-SCH under overload conditions, as shown in Figure~\ref{fig:mpv_sch}. 

Speculative outputs are validated immediately on arrival of the current frame's ISP motion vectors (MV arrival), not at DC completion. If a critical-region delta exceeds a threshold, we perform a local re-extrapolation (no re-inference). The main purpose of DC frames is to realign the application-level result state off the critical path.

\name{}-SPE offers two key benefits. First, it hides latency from downstream tasks that rely on perception outputs. Second, it supports energy-efficient speculation. Since extrapolation is computationally lightweight and downstream tasks such as planning are typically less intensive than perception~\cite{linArchitecturalImplicationsAutonomous2018}, the energy cost of failed speculations remains small.

\revised{\name{}-SPE produces a speculative output one frame ahead and releases it only after the arriving frame's motion vectors confirm the prediction. If the check fails, \name{} discards only this lightweight motion update, never a neural-network inference. It recomputes the output from the measured motion vectors of the arriving frame, the non-speculative \name{}-SCH path, and re-issues the downstream task. This recovery takes milliseconds and uses no neural accelerator. Because nothing is released before confirmation, a downstream task always receives a corrected input and has nothing to roll back. In practice, the erroneous prediction of motion vectors is infrequent and inexpensive.}

\subsection{Reducing Frontend Energy Consumption}\label{sec:frontend_scaling}

\name{}'s motion vector predictions unlock an additional system-level opportunity: enabling adaptive control of the vision frontend to reduce sensing energy. While \name{}-SCH and \name{}-SPE focus on backend efficiency, the image sensor and ISP continue to consume a large share of system energy, especially in Slack regimes where the frontend dominates energy use, accounting for up to 50\% of total consumption~\cite{zhuEuphratesAlgorithmSoCCodesign2018,ganLowLatencyProactiveContinuous2020}.

\name{} enables CV SoCs to proactively anticipate scene dynamics using motion vector predictions and dynamically adjust sensor sampling before the next frame arrives. These predictions can guide energy-aware policies that reduce sensor sampling rate when high fidelity is not required.

We implement a proof-of-concept temporal scaling policy that conserves energy by skipping sensor captures when the predicted motion is low, but only when the application explicitly enables this behavior based on contextual and application-specific insights. For example, an AR headset may activate this policy when the user is focused on a static virtual scene or operating in a low-dynamics environment. Additionally, many CV systems include multiple cameras with distinct roles. In such cases, the system can selectively apply temporal scaling to non-critical sensors, reducing energy consumption while preserving full fidelity on essential visual streams.

In our design, the system enforces temporal consistency by always capturing one frame and conditionally triggering the next based on predicted motion. This alternating pattern creates a controlled ``blinking'' effect during low-motion periods, maintaining a steady flow of perception outputs while reducing sensing activity. For each conditionally triggered frame, the system evaluates the maximum predicted motion vector magnitude $M_{\text{pred}}$ and maps it to a triggering probability:

\[
P_{\text{trigger}} = \min\left(1, \frac{M_{\text{pred}}}{T_{\text{blink}}}\right)
\]

Here, $T_{\text{blink}}$ is a configurable threshold that defines the sensitivity of the policy. This linear mapping produces smooth capture decisions and avoids abrupt transitions that could cause temporal aliasing. When the system skips a frame, it continues operating by using MV predictions to extrapolate perception results, maintaining responsiveness without new visual input. As motion increases, the likelihood of capturing the next frame grows accordingly.

Although more sophisticated sensing strategies could offer further gains, our integration shows that motion predictions provide meaningful utility beyond extrapolation and speculation and serve as a lightweight control signal for adaptive and energy-efficient vision sensing.

%% file: tex/4_arch_proposal.tex
\section{Architectural Support}\label{sec:arch_proposal}

\begin{figure}[b]
\centering
\includegraphics[width=1\linewidth]{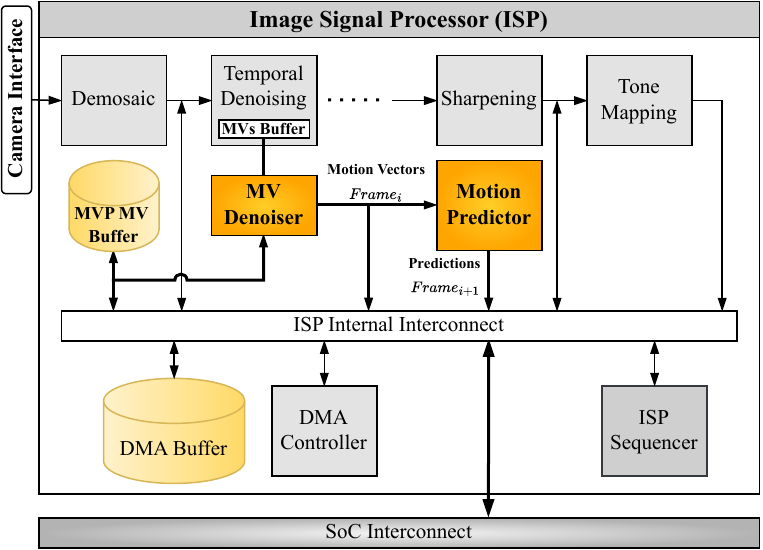}
\caption{Placement of the proposed \name{} components within the ISP pipeline.}
\label{fig:isp_pipeline}
\end{figure}

\name{} augments the ISP with two specialized hardware units, the \textit{MV Denoiser} and the \textit{Motion Predictor}, which are integrated into the pipeline and operate in parallel with existing stages in a streaming fashion, as shown in Figure~\ref{fig:isp_pipeline}. \revised{The MV Denoiser cleans the raw block motion field from the temporal-denoise stage so that it is spatially coherent and temporally stable, and the Motion Predictor forecasts each tile's next-frame motion vectors from their recent history.}

Modern ISPs commonly perform block-based motion estimation as part of the Temporal Denoising (TD) stage. This process partitions each frame into fixed-size blocks, typically $16\times16$ pixels, and searches for the most similar block in a previous frame. The search is performed within a window using a matching cost function such as the Sum of Absolute Differences (SAD). The block with the lowest SAD becomes the best match, and the resulting spatial offset defines the motion vector capturing the block's movement between frames.

The TD stage continuously generates motion vectors for each block and stores them in internal ISP buffers for tasks such as denoising or stabilization. \name{}  intercepts this internal stream and reroutes these motion vectors into our proposed datapath. However, raw motion vectors are often noisy due to occlusions, low-texture regions, repetitive patterns, or lighting variations. Thus, the first step in \name{}'s datapath is to spatially and temporally filter the raw vectors to improve their reliability.

\subsection{MV Denoiser}

\begin{figure}[t]
\centering
\includegraphics[width=0.9\linewidth]{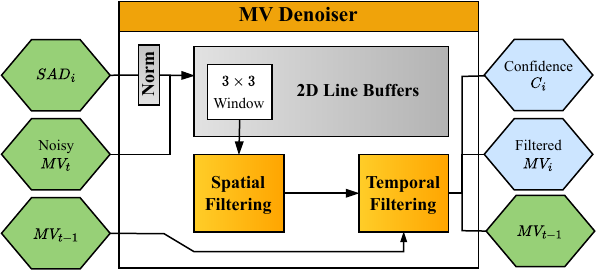}
\caption{The MV Denoiser filters raw motion vectors produced by the temporal denoising stage, enhancing spatial coherence and temporal stability.}
\label{fig:mv_denoiser}
\end{figure}

The MV Denoiser (Figure~\ref{fig:mv_denoiser}) processes the stream of motion vectors using a hardware-friendly line-buffer design, a common structure in image processing pipelines~\cite{ujjainkarImaGenGeneralFramework2023}. It applies two stages of filtering: spatial smoothing followed by temporal fusion. The unit stores the final filtered MVs in a compact buffer (8 bits per component, totaling approximately 16\,KB for 1080p resolution with $16 \times 16$ blocks) exposed to the ISP (see Figure~\ref{fig:isp_pipeline}).

The first stage of the MV Denoiser applies a 3$\times$3 spatial averaging filter to suppress local noise while preserving coherent motion patterns. This is based on the observation that motion is typically smooth over small spatial neighborhoods, where nearby blocks exhibit similar displacement.

To improve robustness, the averaging is weighted by a reliability score, or \textit{confidence}, assigned to each motion vector, as used in prior work~\cite{zhuEuphratesAlgorithmSoCCodesign2018}. This confidence reflects the quality of the block match during motion estimation and is derived from the SAD computed by the ISP. A lower SAD value indicates a better match and therefore a higher confidence.

For each motion vector centered at block $(x, y)$, the filtered output $\vec{MV}_f(x, y)$ is computed as:

\begin{equation}
\vec{MV}_f(x, y) = \frac{1}{Z(x, y)} \sum_{i,j=-1}^{1} c(x{+}i, y{+}j)\,\vec{MV}(x{+}i, y{+}j)
\end{equation}

where $c(x+i, y+j)$ is the confidence score for each neighboring block, and $Z(x, y)$ is the normalization factor:

\begin{equation}
Z(x, y) = \sum_{i=-1}^{1} \sum_{j=-1}^{1} c(x+i, y+j)
\end{equation}

The confidence $c(x, y)$ is computed from the SAD value as:

\begin{equation}
c(x, y) = 1 - \frac{\text{SAD}(x, y)}{L^2 \cdot 255}
\end{equation}

where $\text{SAD}(x, y)$ is the block matching cost and $L$ is the block size (for example, $L = 16$).

After spatial filtering, the denoised vector is fused with its counterpart from the previous frame using a confidence-aware exponential moving average (EMA):

\begin{equation}
\alpha(x, y) = \max(0.5, c(x, y))
\end{equation}
\begin{equation}
\begin{split}
\vec{MV}_t(x, y) = {} & \alpha(x, y) \cdot \vec{MV}_f(x, y) \\
& + (1 - \alpha(x, y)) \cdot \vec{MV}_{t-1}(x, y)
\end{split}
\end{equation}

The hardware footprint of the MV Denoiser is minimal. It requires a 3-line buffer to implement the 3$\times$3 sliding window and a small number of fixed-point multipliers and adders.

\subsection{Motion Predictor}

\begin{figure}[b]
\centering
\includegraphics[width=0.9\linewidth]{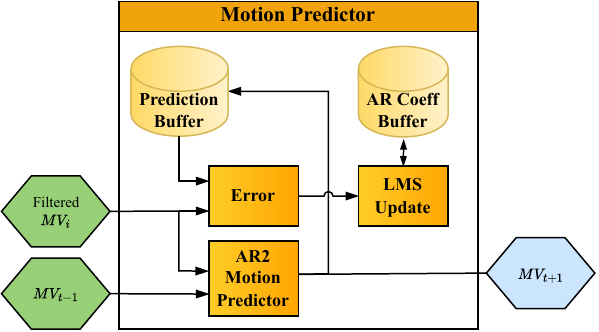}
\caption{\name{} Motion Predictor. Predicts next-frame motion vectors using a tile-wise AR(2) model.}
\label{fig:motion_predictor}
\end{figure}

The design of the Motion Predictor (Figure~\ref{fig:motion_predictor}) is guided by the goal of enabling low-cost, localized, and adaptive motion extrapolation with minimal hardware overhead. Motion in video tends to be locally smooth and temporally correlated: small frame regions typically exhibit coherent, predictable movement over short time windows. Rather than relying on global motion models or expensive learning-based predictors, we adopt a lightweight, interpretable approach based on second-order autoregressive (AR(2)) filtering.

To capture spatial variation in motion dynamics, the motion field is divided into a grid of non-overlapping tiles, each comprising a fixed number of motion blocks (e.g., $4 \times 4$ blocks per tile, with $16 \times 16$ pixels per block). Each tile $k$ maintains an independent AR(2) predictor, parameterized by a pair of 2D coefficient vectors $\vec{a}_1(k)$ and $\vec{a}_2(k)$ that separately scale the horizontal and vertical motion components. This allows each region to learn independently and adapt to local motion behavior.

For each block $(x, y)$ belonging to tile $k$, the predictor estimates the next-frame motion vector using the AR(2) model:

\begin{equation}
\begin{split}
\hat{\vec{MV}}_{t+1}(x, y) = {} & \vec{a}_1(k) \cdot \vec{MV}_t(x, y) \\
& + \vec{a}_2(k) \cdot \vec{MV}_{t-1}(x, y)
\end{split}
\end{equation}

The coefficients $\vec{a}_1(k)$ and $\vec{a}_2(k)$ are 2D vectors that scale the horizontal and vertical components of motion independently. These are updated online using a lightweight Least Mean Squares (LMS) adaptation rule. As new motion vectors arrive from the ISP, the predictor immediately compares each observed vector $\vec{MV}_t(x, y)$ to the previously predicted value $\hat{\vec{MV}}_t(x, y)$ and computes the prediction error $\vec{e}_t(x, y)$, defined as the difference between the observed and predicted vectors. This error is then used to update the AR coefficients for tile $k$:

\begin{equation}
\vec{a}_i(k) \leftarrow \vec{a}_i(k) + \eta \cdot \left( \vec{e}_t(x, y) \cdot \vec{MV}_{t - i + 1}(x, y) \right)
\end{equation}

where $i \in \{1, 2\}$, $\eta$ is a small learning rate, and $\cdot$ denotes a product per component (element-wise). \revised{In hardware, the predictor uses the velocity form of this model, $\vec{MV}_t + \vec{\beta}(k)\cdot(\vec{MV}_t - \vec{MV}_{t-1})$, with one coefficient $\vec{\beta}$ per tile between 0 and 1. The update step is divided by the input power, so large motion vectors cannot make it unstable, and $\vec{\beta}$ is kept within its bounds. A tile that learns a poor $\vec{\beta}$ does no harm. Its prediction fails the validation check, that frame uses the measured motion vectors instead, and only the speedup is lost.}

The predictor maintains buffers for previous predictions (to compute error) and a small set of AR coefficients per tile (2\,KB for 1080p frames). It writes the predicted vectors directly to the ISP's DMA buffer, which the DMA engine then transfers to the SoC's main memory.

%% file: tex/5_methodology.tex
\section{Methodology}
\label{sec:methodology}

We evaluate Balanced and Overload regimes under identical sensing and compute budgets, and report mAP, energy, and P99 tail latency. Energy is counted per produced perception output, so a pipeline that drops frames is charged only for the frames it processes, and latency spans the full path from sensing to downstream-task completion. Real-time control flow is modeled explicitly: bounded queue, explicit drop policy, and validation on MV arrival. SoC timing uses gem5 with DRAM/NPU models, and the ISP is validated with a cycle-accurate, cascaded, line-buffered ready/valid RTL micro-pipeline calibrated to 1080p@30\,FPS. \name{} adds two streaming stages on the motion path: MV Denoiser ($3\times3$+EMA) and tile-wise AR(2) Motion Predictor.

We evaluate \name{} on object detection, a compute-intensive perception task central to real-time vision systems. Using YOLOv8~\cite{Jocher_Ultralytics_YOLO_2023} pretrained on COCO~\cite{linMicrosoftCOCO2014}, we run experiments on the camera-based KITTI benchmarks~\cite{KITTIVisionBenchmark}, an established autonomous-driving dataset with urban, residential, and highway scenes covering a range of activity levels. This variability allows us to assess \name{} under both low- and high-complexity conditions. KITTI is captured at 10\,Hz, so we upsample it to 30\,FPS with RIFE~\cite{huangRIFE2022} and keep the original frames as every third frame. We score accuracy on the original frames and their labels only. On extrapolated frames, we translate each box by the mean motion vector of the blocks it covers. We also rescale the box using the difference between the mean vectors of its left and right halves, and likewise for its top and bottom halves. We use an IoU threshold of 0.5 and report mean Average Precision (mAP@0.5) across all classes.

\revised{We additionally evaluate \name{} on multi-object tracking, a common and safety-critical workload in continuous-vision systems. We use ByteTrack}~\cite{bytetrack}\revised{, a state-of-the-art lightweight tracker for mobile and edge platforms, on the standard KITTI tracking and MOT17}~\cite{mot17}\revised{ benchmarks, the latter being a crowded non-driving pedestrian dataset used without fine-tuning, and we report the Higher Order Tracking Accuracy (HOTA). We also evaluate the transformer detector RT-DETR}~\cite{rtdetr}\revised{ alongside YOLOv8.}

To analyze system-level behavior, we build a real-time simulator that models the complete vision pipeline. Instead of re-executing each component, we combine measured performance and power profiles with validated architectural models, enabling fast and accurate evaluation of pipeline-level effects such as scheduling decisions, stalls, and frame drops.

We model an AR1335 image sensor~\cite{ar1335} operating at 30 FPS, using power values from vendor datasheets. The baseline SoC includes an ISP, CPU, and NPU. We use Jetson TX2 specifications to model the ISP,  and validate the integration of \name{} with a cycle-accurate RTL pipeline. This ensures that motion vector denoising and prediction modules do not introduce processing stalls (Figure~\ref{fig:isp_pipeline}). Additionally, we extract motion vectors from video sequences using OpenCV \revised{sparse pyramidal Lucas-Kanade feature tracking}~\cite{bradskiOpenCVLibrary2000}\revised{, which performs local windowed matching with per-block validity, like the motion estimator inside an ISP, so it is representative of an ISP-exported motion-vector stream. To confirm this, we compared the sparse-LK fields against H.264 encoder block motion estimation, the same SAD block-matching that ISP temporal-denoise stages run. The two agree to a median endpoint error near one pixel across the KITTI sequences. At the task level, warped-detection recall differs by at most two points between the two sources, with the encoder source equal or slightly better.}

We run the extrapolation and scheduling logic on a gem5~\cite{lowe-powerGem5SimulatorVersion2020}-modeled ARM Cortex-A72 CPU. We simulate DRAM behavior using DRAMsim3~\cite{liDRAMsim3CycleAccurateThermalCapable2020}. For vision inference, we use ScaleSimv2~\cite{samajdarSystematicMethodologyCharacterizing2020} to model an RTL-validated NPU with a $24\times24$ MAC array at 1\,GHz and 1.5\,MB of double-buffered SRAM. \revised{This single-NPU budget is the typical mobile-SoC configuration, so our comparisons against speculative methods assume one shared backend.}

We evaluate two compute regimes. In the Balanced regime, we use YOLOv8s (28.6B FLOPs), which fits within the NPU's compute capacity. In the Overload regime, we use YOLOv8m (78.9B FLOPs), which exceeds the NPU budget and sustains only 14.7 FPS. \revised{We report all reductions against the conventional serialized pipeline of} Figure~\ref{fig:regimes_cv}a\revised{, and we compare with the following techniques:}

\begin{itemize}
\item \textbf{CME}: Classical motion extrapolation mimicking Euphrates \cite{zhuEuphratesAlgorithmSoCCodesign2018} with motion extrapolation performed on the CPU and detection on the NPU.
\item \textbf{PDSPEC}: Pixel-domain speculation based on PVF~\cite{ganLowLatencyProactiveContinuous2020}, evaluated under the same constraints as the other techniques, with no frontend shutdown and a single NPU used for both real and speculative frames.
\item \textbf{\name{}-SCH-A}: Our scheduler that decouples extrapolation from full-frame processing and applies adaptive DCFD regulation (see Section~\ref{sec:DCFD}).
\item \textbf{\name{}-SPE-A}: Extends \name{}-SCH-A with the proposed speculative motion extrapolation.
\item \textbf{\name{}-ORACLE}: An ideal variant of \name{}-SPE-A that assumes perfect predictions to represent the upper bound of speculation performance.
\end{itemize}

%% file: tex/6_results.tex
\section{Experimental Results}\label{sec:results}

This section presents a detailed evaluation of \name{} in the context of object detection, a representative computer vision workload. We analyze its performance across the compute regimes defined in Section~\ref{sec:regimes_cv} and compare it against state-of-the-art techniques described in Section~\ref{sec:prior_approaches}.

We denote configurations as \name{}-SCH-A (adaptive scheduling), \name{}-SPE-A (adaptive speculation), and \name{}-ORACLE (perfect prediction upper bound), as defined in Section~\ref{sec:methodology}. Across all configurations, we fix the Max Speculation Distance (MSD) to 1, meaning the system speculates only one frame ahead (i.e., predicting frame $i{+}1$ during the processing of frame $i$). To better reflect latency-sensitive real-time workloads, we introduce a 10ms downstream payload task, representing the worst-case execution time (WCET) of typical perception consumers such as planning or control~\cite{linArchitecturalImplicationsAutonomous2018}.

To preserve accurate perception, we set a target maximum degradation of 1.5 mAP points, consistent with values used in prior studies~\cite{zhuEuphratesAlgorithmSoCCodesign2018,ganLowLatencyProactiveContinuous2020}. For context, this target remains well below the accuracy drop from downsizing the backbone: moving from YOLOv8s to the smaller YOLOv8n (not evaluated here) reduces mAP by more than 7 points. In contrast, \name{} sustains competitive accuracy while delivering higher throughput and lower latency, as we show in the following sections.

\subsection{Accuracy Analysis}\label{subsec:results_accuracy}

\begin{figure}[t]
    \centering
    \includegraphics[width=0.9\linewidth]{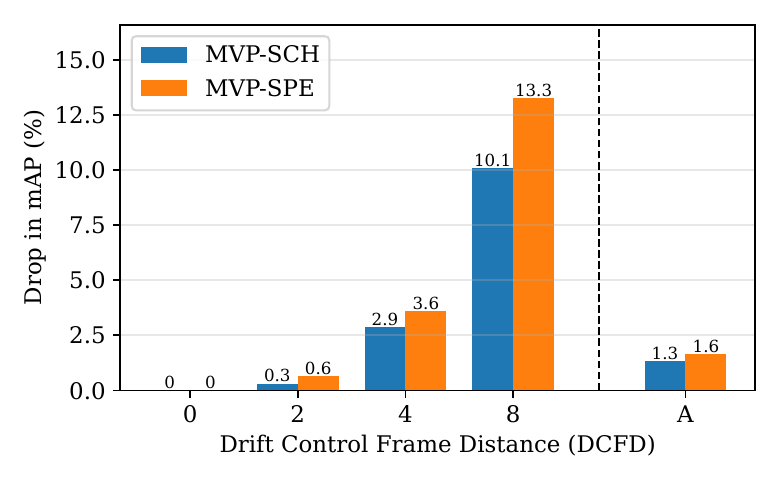}
    \caption{Impact of different \name{} configurations on YOLOv8 accuracy. The plot shows mAP drop (\%) relative to the baseline as DCFD increases. ``A'' denotes the adaptive \name{} configuration.}
    \label{fig:mvp_yolo_accuracy}
\end{figure}

Figure~\ref{fig:mvp_yolo_accuracy} shows the impact of different \name{} configurations on semantic accuracy, measured as the percentage drop in mAP across varying values of the DCFD parameter, with a MED of DCFD $+$ 1. The drop is measured against the same detector run on every frame, so it captures the error introduced by extrapolation and not the error of the detector itself. We report it for YOLOv8m as the more demanding case.

Across all DCFD values, both \name{}-SCH and \name{}-SPE maintain high accuracy at short horizons. At DCFD$=$0, where every frame is corrected, the output equals the baseline. The drop stays below 1\% up to DCFD$=$2 and reaches 2.9\% (\name{}-SCH) and 3.6\% (\name{}-SPE) at DCFD$=$4. At DCFD$=$8, where each output is extrapolated up to 267\,ms from the last corrected frame, the drop grows to 10\% and 13\%. \name{}-SPE stays within one point of \name{}-SCH up to DCFD$=$4, since speculation adds only the prediction error of a single frame of motion on top of the extrapolation drift. The gap widens with the horizon as these errors compound. The adaptive configuration, discussed next, is the operating point we recommend, and it keeps the cost between 1.3\% and 1.6\%.

The adaptive configuration, shown as the ``A'' point in Figure~\ref{fig:mvp_yolo_accuracy}, adjusts the DCFD with the deviation threshold of Section~\ref{sec:DCFD}, set so that the resulting drop stays within the 1.5 mAP target, and a maximum DCFD$=$4. It achieves accuracy close to the target while adapting to the scene. In our experiments, the scheduler stays at DCFD$=$4 most of the time and shortens the distance on scene changes, reducing the frequency of DC corrections without exceeding the accuracy bound.

\revised{We also evaluate \name{} on the tracking and transformer-detector settings of} Section~\ref{sec:methodology}\revised{, summarized in} Table~\ref{tab:workloads}\revised{. On ByteTrack tracking over the KITTI sequences, the extrapolated detections track at a few-point HOTA cost, concentrated in identity association instead of detection, and MOT17 shows the same behavior in a crowded non-driving domain. The transformer detector RT-DETR runs unchanged and incurs a 2.6\% mAP cost at DCFD$=$4, which the adaptive scheduler reduces to 1.6\%. The latency and energy reductions carry over to these workloads, since they come from the scheduling and from avoiding detector inference, both independent of the task. The fixed DCFD$=$4 saves more energy than the adaptive configuration because it runs fewer correction frames, and the heavier RT-DETR detector saves slightly more energy.}

\begin{table}[t]
\centering
\caption{\name{} on additional workloads (DCFD$=$4).}
\label{tab:workloads}
\begin{tabular}{@{}lll@{}}
\toprule
\revised{\mbox{Workload (detector)}} & \revised{\mbox{Accuracy}} & \revised{\mbox{Energy}} \\
\midrule
\revised{\mbox{KITTI track (YOLOv8m)}} & \revised{\mbox{HOTA 0.483$\rightarrow$0.463}} & \revised{\mbox{58\%}} \\
\revised{\mbox{MOT17 (YOLOv8m)}} & \revised{\mbox{HOTA 0.411$\rightarrow$0.389}} & \revised{\mbox{58\%}} \\
\revised{\mbox{KITTI (RT-DETR)}} & \revised{\mbox{mAP $-$2.6\%}} & \revised{\mbox{63\%}} \\
\bottomrule
\end{tabular}
\end{table}

\subsection{Latency and Energy in the Balanced Regime}\label{subsec:lat_analysis}

\begin{figure}
    \centering
    \includegraphics[width=0.9\linewidth]{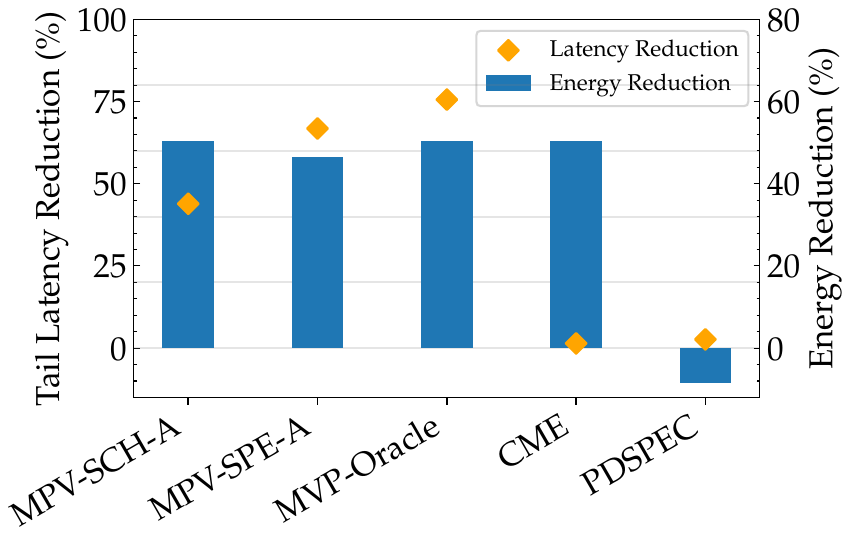}
    \caption{Tail latency reduction and energy consumption reduction under Balanced regime conditions for different \name{} configurations, compared with the most relevant state-of-the-art alternatives.}
    \label{fig:mvp_balanced_latency}
\end{figure}

Figure~\ref{fig:mvp_balanced_latency} presents the tail latency and energy savings achieved by various \name{} configurations, alongside the most competitive state-of-the-art baselines, under a Balanced regime.

Both \name{}-SCH-A and \name{}-SPE-A outperform prior techniques by using motion extrapolation as the default processing path while executing drift correction in parallel. Among them, \name{}-SPE-A achieves the lowest tail latency (around 22\% extra reduction) by speculatively producing results one frame ahead and triggering downstream tasks early. The \name{}-ORACLE configuration demonstrates the upper bound of speculation performance. It shows that, in the absence of prediction errors, tail latency could be reduced by a further 8.78\%, and that failed predictions cost \name{}-SPE-A 4.82\% in energy relative to this bound.

When compared to CME, which offers similar energy savings, \name{}-SPE-A delivers markedly lower tail latency. This difference arises because CME periodically processes entire frames without extrapolation, forcing the rest of the pipeline to wait for results. 

\name{} also outperforms speculative methods like PDSPEC, which operate in the pixel domain and require full inference for each speculative frame. \name{} reduces tail latency with much lower energy overhead. PDSPEC speculatively processes the next frame during NPU idle periods, and if the prediction closely matches the actual frame, it can continue forward without re-execution. However, as speculation continues over multiple frames, the predicted frames diverge from the actual ones. In our experiments, this divergence typically leads to mispredictions after 4–5 frames, triggering backend reprocessing. These limit the achievable tail latency reduction and introduce around 10\% additional energy consumption.

\subsection{Worst-case Misprediction Behavior}\label{subsec:worst_case_misprediction}

To bound the behavior of \name{} under adverse conditions, we consider a worst-case scenario in which all speculative predictions fail validation (i.e., every predicted motion-vector field for frame $i{+}1$ is rejected). In this limit, \name{}-SPE degenerates to \name{}-SCH: speculative outputs are discarded and the system falls back to the non-speculative motion-extrapolation path anchored at the last drift-corrected frame. In practice, the observed misprediction rate across KITTI sequences averages 8--12\% of frames, concentrated in scenes with abrupt motion changes. Tail latency, throughput, and accuracy therefore match those of \name{}-SCH up to the small fixed cost of running the predictor and validation (since it is done in hardware at the ISP). In contrast, PDSPEC schemes (e.g., PVF~\cite{ganLowLatencyProactiveContinuous2020}) must generate full frames and run full backend inference for each speculative frame. Under 100\% misprediction, all speculative results are discarded and every frame must be reprocessed on the actual input, doubling backend work and pushing the system deeper into the Overload regime with higher energy.

\name{}-SCH already improves over CME because drift correction is decoupled from the critical path: full backend inference runs asynchronously in the background, while extrapolated results remain non-blocking for all incoming frames. However, both CME and \name{} must reduce the extrapolation distance to zero in the worst case. In that extreme scenario, \name{} does not provide benefits (similarly to CME) but does not hurt latency. It only increases the energy by a tiny amount to support the MV prediction. Taken together with the \name{}-ORACLE configuration, which captures the ideal case of perfect predictions, this analysis shows that \name{}'s envelope ranges from an improved extrapolation-only scheduler in the worst case to a strictly better latency–energy trade-off than prior work when predictions are accurate (Figure~\ref{fig:mpv_overload_throughput}).

\revised{A newly appearing object is different from a tracked one. Extrapolation moves boxes the detector has already found, so it cannot introduce an object the detector has not yet seen. A new object therefore waits for the next drift-correction frame, which runs the full detector. This onset delay is the information-propagation latency of} Section~\ref{sec:regimes_cv}\revised{, a latency effect rather than a loss in per-frame accuracy. At DCFD$=$4 and 30\,FPS, this worst case is about 235\,ms: up to four 33\,ms frame periods until the next correction, one 68.5\,ms inference, and one 33\,ms frame to propagate the result. Every frame-based system pays the same floor of one frame plus one inference for new content, and under the Overload regime the baseline sustains only 14.7\,FPS, so its backlog pushes its own worst case higher still. \name{} is therefore no worse than the baseline under overload. Finally, the system designer has the final word in bounding this worst case by lowering the DCFD according to the specific application.}

\subsection{Throughput and Energy in the Overload Regime}\label{subsec:throughput_analysis}

\begin{figure}
    \centering
    \includegraphics[width=0.9\linewidth]{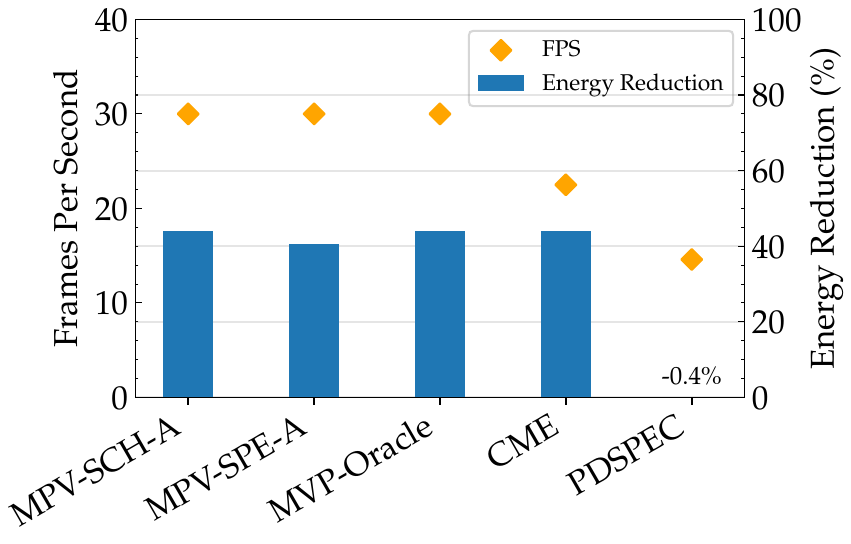}
    \caption{FPS throughput and energy reduction under Overload regime for different \name{} configurations and the most relevant state-of-the-art alternatives.}
    \label{fig:mpv_overload_throughput}
\end{figure}

Figure~\ref{fig:mpv_overload_throughput} compares the throughput and energy efficiency of \name{} and baseline approaches in the Overload regime, where backend latency exceeds the frame interval and causes frame drops. Since \name{} still relies on motion extrapolation, tail-latency trends resemble those in the Balanced regime; the key difference here is each method's ability to sustain throughput under persistent computational pressure.

In \name{}, both \name{}-SCH-A and \name{}-SPE-A maintain low latency by design, as motion extrapolation remains the default processing path. However, operating under overload requires a higher MED (3 in this experiment) to account for longer backend delays. Despite this, the predictor remains accurate enough to preserve semantic quality, and the system continues to provide timely results without blocking.

PDSPEC fails to improve throughput in this regime. Because we assume only one backend execution unit, there is no idle time available for speculative execution. As a result, the system behaves like the baseline, sustaining a throughput of only 14.7 FPS. The additional logic for speculative control introduces a slight energy overhead without delivering performance gains.

CME improves throughput over the baseline by skipping some full-frame processing steps, reaching approximately 23~FPS versus the baseline's 14.7~FPS. However, it still experiences frame drops because the object detector takes longer than two frames to process each input. CME also periodically processes full frames without extrapolation, introducing blocking delays that the rest of the pipeline must absorb. As a result, CME cannot match \name{}'s sustained throughput and fails to maintain real-time responsiveness under the flow-control policy described in Section~\ref{sec:regimes_cv}.

In contrast, \name{} configurations sustain higher throughput and consistently low latency, even under constrained compute. While this requires extrapolating over longer distances, it remains preferable to dropping frames, as it allows the system to continue delivering timely and meaningful outputs. The energy reduction in Figure~\ref{fig:mpv_overload_throughput} is per output, as defined in Section~\ref{sec:methodology}. \name{} delivers twice the outputs of the baseline at that cost.

\subsection{Frontend Scaling Analysis}\label{subsec:scalin_analysis}

We evaluate the potential of \name{}'s motion prediction to modulate temporal sampling, following Section~\ref{sec:frontend_scaling}. This optional mechanism reduces sensor activity based on predicted motion and is suitable for applications that do not require guaranteed visual feedback.

We evaluate three $T_{\text{blink}}$ thresholds (Section~\ref{sec:frontend_scaling}): conservative ($T_{\text{blink}}=2$\,px), moderate ($T_{\text{blink}}=4$\,px), and aggressive ($T_{\text{blink}}=8$\,px).

Assuming an energy-proportional sensor~\cite{likamwaEnergyProportionalImage2013}, the conservative and moderate settings achieve 5.4\% and 13.8\% frontend energy savings, respectively, while staying within the 1.5 mAP target. The aggressive mode cuts nearly 50\% of frontend energy but incurs a 6.2\% mAP drop. Future systems with higher frame rates could tolerate more aggressive scaling with smaller accuracy tradeoffs.

\subsection{\texorpdfstring{\revised{Area, Timing, and Power}}{Area, Timing, and Power}}\label{subsec:results_area_power}

We evaluate the area and power impact of \name{} using the methodology in Section~\ref{sec:methodology} targeting support for Full HD resolution on a 12\,nm-class mobile process. The additional hardware required by the new \name{} ISP augmentations incurs a minimal area cost of 0.11\,mm$^2$, primarily due to approximately 32\,KB of on-chip SRAM for the motion-vector and prediction buffers, 2\,KB for the autoregressive model coefficients, and the compact 3$\times$3 line buffer used by the MV Denoiser. For context, a typical mobile ISP occupies 5--15\,mm$^2$ depending on feature set and process node. \name{}'s addition represents less than 2\% of even a modest ISP, and a negligible fraction of full mobile SoCs~\cite{A14BionicApple,Nvidia_XavierHotchips2018}. %

To obtain measured timing and power and a logic-versus-memory area breakdown, we implemented the MV Denoiser and the AR(2) Motion Predictor in synthesizable RTL, checked their outputs against a fixed-point software model of Section~\ref{sec:arch_proposal}, and synthesized them with an open flow (Yosys~\cite{wolfYosysOpenSYnthesis}, OpenROAD~\cite{openroad}, and OpenSTA~\cite{opensta}) in the ASAP7 7\,nm standard-cell library~\cite{asap7}. Table~\ref{tab:synth} reports the result. The design closes timing at 708\,MHz, with the critical path in the denoiser exponential-moving-average multiply-add, and it draws 0.59\,mW of dynamic power and 3.4\,$\mu$W of leakage. The motion-vector SRAM dominates the cell area at about 65\%, and the denoiser and predictor logic accounts for the remaining 35\%. These results give the measured timing, power, and the logic-versus-memory composition. We report the absolute area as the 0.11\,mm$^2$ estimate from our system-level model at the 12\,nm-class node, a conservative figure relative to scaling the synthesized design to that node following Stillmaker and Baas~\cite{stillmaker}, and the synthesis provides its logic-versus-memory composition. Both units operate as non-stalling streaming stages that sustain over 45$\times$ the required block rate, so they integrate on the motion path without back-pressuring the existing ISP.

\begin{table}[t]
\centering
\caption{Synthesis results for the \name{} ISP units.}
\label{tab:synth}
\begin{tabular}{ll}
\toprule
Parameter & Value \\
\midrule
Standard-cell library & ASAP7 RVT, TT \\
Tools & Yosys, OpenROAD, OpenSTA \\
Max.\ frequency & 708\,MHz \\
Critical path & denoiser EMA multiply-add \\
Dynamic power (50\,MHz) & 0.59\,mW \\
Leakage power & 3.4\,$\mu$W \\
Logic / SRAM area & 35\% / 65\% \\
\bottomrule
\end{tabular}
\end{table}

%% file: tex/7_related_work.tex
\section{Related Work}\label{sec:related_work}

While Section~\ref{sec:motivation} already discusses two of the most relevant prior works, Euphrates~\cite{zhuEuphratesAlgorithmSoCCodesign2018} and PVF~\cite{ganLowLatencyProactiveContinuous2020}, we now review broader research efforts that also aim to exploit temporal redundancy, reduce processing overhead, or improve coordination between hardware and software in CV pipelines.

Optical flow prediction estimates how each pixel in a video frame will move in future time steps, producing dense and detailed motion fields. Prior work has explored deep learning models that predict optical flow from sequences of past frames~\cite{dongMemFlowOpticalFlow2024}, from past motion vectors~\cite{ciamarraForecastingFutureInstance2022}, or even from a single image~\cite{walkerDenseOpticalFlow2015, argawOpticalFlowEstimation2021}. These methods often use architectures such as ConvLSTMs, UNets, or variational autoencoders to model motion and uncertainty over time. While they are effective at fine-grained prediction, their computational cost makes them difficult to deploy in real-time on embedded platforms. In contrast, our method predicts motion in a coarse block-based domain using lightweight autoregressive models fed by motion vectors already available from the ISP. This approach is simple, efficient, and suitable for integration in real-world mobile systems. We leave it to future work to explore whether these sophisticated optical flow prediction models can be adapted for embedded use. We expect ongoing research to keep improving motion prediction, which, as this work shows, benefits future CV SoCs.

EVA$^2$~\cite{bucklerEVAExploitingTemporal2018} introduces activation motion compensation (AMC), a method inspired by video compression, where motion estimation is used to incrementally update intermediate CNN activations rather than recomputing them for every frame. This reduces redundant computation by only refreshing activations when significant visual changes are detected, achieving substantial energy savings with minimal impact on accuracy. EVA$^2$ AMC allows \name{} to apply extrapolation/speculation to CNNs.

Recent work has increasingly highlighted the ISP's role in reducing redundancy early in the CV pipeline. Rhythmic Pixel Regions (RPR)~\cite{kodukulaRhythmicPixelRegions2021} is a mechanism that allows the backend to dictate spatial and temporal resolution policies for image regions. $\delta$LTA~\cite{tarancoDLTADecouplingCamera2023} enables region-level frame skipping by filtering out temporally redundant image regions before backend execution. IRIS~\cite{tarancoIRISUnleashingISPSoftware2025} partially downscales static regions using saliency cues derived from motion vectors and edge density. These techniques can further optimize \name{}'s drift-correction frames by reducing their cost, while \name{} can guide their activation using proactive motion prediction. By predicting scene dynamics before frame arrival, \name{} enables more efficient frontend-backend coordination.

Finally, several approaches address redundancy directly at the point of capture through adaptive frontend control. Glimpse~\cite{naderipariziGlimpseProgrammableEarlyDiscard2017} proposes a low-power vision pipeline that uses lightweight motion cues to trigger full-resolution capture only when necessary, discarding uninformative frames early to reduce sensing and processing overhead. This enables continuous vision on power-constrained devices by avoiding costly per-frame analysis. Earlier work by LiKamWa et al.~\cite{EnergyCharacterizationOptimization} demonstrated that modern CMOS sensors already support underutilized energy-proportional features, such as clock scaling and low-power standby modes, which substantially reduce power at lower frame rates. We show that predicting future motion vectors can serve as an effective control signal for the frontend, complementing prior techniques to support better proactive sensor scaling.

%% file: tex/8_conclusions.tex
\section{Conclusions}\label{sec:conclusions}

This paper presented \name{}, a motion-predictive speculative vision pipeline that tackles the latency and energy challenges of real-time continuous vision. By operating entirely in the motion domain, \name{} uses a decoupled scheduling model that delivers immediate perception via motion extrapolation while relegating full backend processing to a non-blocking background task. Building on this, motion-based speculation and a lightweight ISP-embedded predictor enable proactive, low-cost execution and optional frontend scaling. Our evaluation shows that \name{} substantially reduces tail latency and energy consumption at a small accuracy cost, indicating that motion-aware pipelines provide a practical path to new performance–efficiency operating points for mobile and embedded CV systems.